\documentclass{Interspeech}

\interspeechcameraready

\title{Multimodal Fusion with Semi-Supervised Learning Minimizes Annotation Quantity for Modeling Videoconference Conversation Experience}

\author[affiliation={1}]{Andrew}{Chang}
\author[affiliation={1}]{Chenkai}{Hu$^*$}
\author[affiliation={1}]{Ji}{Qi$^*$}
\author[affiliation={1}]{Zhuojian}{Wei$^*$}
\author[affiliation={1}]{Kexin}{Zhang$^*$}
\author[affiliation={1}]{Viswadruth}{Akkaraju}
\author[affiliation={1,2}]{David}{Poeppel}
\author[affiliation={1}]{Dustin}{Freeman}

\affiliation{}{New York University}{USA}
\affiliation{}{Max Planck Society}{Germany} 
\email{\{ac8888, ckh326, jq2316, zw2219, kz1267, va2353, dp101, df2581\}@nyu.edu}

\usepackage{comment, color, soul}

\begin{document}

\maketitle

\begin{abstract}

Group conversations over videoconferencing are a complex social behavior. However, the subjective moments of negative experience, where the conversation loses fluidity or enjoyment remain understudied. These moments are infrequent in naturalistic data, and thus training a supervised learning (SL) model requires costly manual data annotation. We applied semi-supervised learning (SSL) to leverage targeted labeled and unlabeled clips for training multimodal (audio, facial, text) deep features to predict non-fluid or unenjoyable moments in holdout videoconference sessions. The modality-fused co-training SSL achieved an ROC-AUC of 0.9 and an F1 score of 0.6, outperforming SL models by up to 4\% with the same amount of labeled data. Remarkably, the best SSL model with just 8\% labeled data matched 96\% of the SL model's full-data performance. This shows an annotation-efficient framework for modeling videoconference experience.

\end{abstract}

\renewcommand{\thefootnote}{\fnsymbol{footnote}} \footnote[0]{$^*$Equal contribution; authors listed alphabetically.}

\section{Introduction}
Videoconferencing has become a common and irreplaceable channel for communication in both professional and casual contexts, especially in the post-COVID-19 era. Although it is an essential medium for communication, it has not been sufficiently studied. One of the major differences is that, compared to in-person contexts, it is not yet as effective for fostering fluid conversation, enjoyable interaction, and psychological intimacy \cite{zhao2023separable, mignault2024perceiving, balters2023virtual}. While the causes of these moments of negative experiences remain unclear and may involve both videoconferencing signals and the human perceptual and cognitive factors, modeling these moments is critical for enhancing user experience and human-computer interactions.

While collecting long recordings of videoconferencing data is relatively easy, most of the time in a session is usually uneventful and provides little value for failure analysis. However, the infrequent negative moments can have a significant impact on the overall user experience \cite{powers2011effect}. This imbalance in both frequency and impact makes training a machine learning model to identify infrequent negative moments challenging. We risk overlooking the potential benefits of leveraging uneventful periods to improve the model while also facing high annotation costs if every moment in a long recording is labeled. 

In this study, by experimenting with multiple SSL approaches to leveraging both labeled/unlabeled targeted and non-targeted clips, we aimed to effectively train a classifier on multimodal (audio, face, text) deep feature embeddings to predict the clips with negative moments on the holdout videoconference sessions. Specifically, we identified the relationship between labeled sample size, SSL and SL model performances, and the contribution of fused and individual modalities.

\section{Prior Works}
To the best of our knowledge, SSL has not been used to model videoconferencing experience. While audio-based SSL has been widely applied in speech emotion recognition, multimodal approaches have only recently emerged \cite{andrade2022survey}. For instance, a study used SSL and a visual-to-audio knowledge transfer approach to improve the robustness of the model \cite{zhang2021combining}. Another study developed an SSL approach that enabled audio-text interactions and successfully modeled speech emotions \cite{lian2022smin}. In a recent competition, a study using audio, video, face and text features to train an SSL model to accurately recognized speech emotion with only 1\% data labeled \cite{qi2024multimodal}. While these successes in a single speaker’s speech audio imply that multimodal SSL is a promising approach for modeling videoconference experience, the phenomena is qualitatively different as videoconference experience is heavily influenced by the interactions and turn-taking behaviors among the participants, which emergent features cannot be simply reduced to the combination of individual speakers.

A few relevant recent works used supervised learning (SL) approaches with multimodal (audio, speech, face and/or body motion) features to model the user experience of videoconferencing \cite{vrzakova2020focused, Bingol2024, chang2025multimodal}. However, we should not assume that the SL approach can be directly generalized to SSL, as the feature representations may have a complex relationship with user experience. For example, overlapping speech from multiple speakers could be perceived as a disruptive interruption or a helpful and enthusiastic backchannel, leading to very different subjective experiences. This potential nonlinear behavior may be modelable using SL, but it could violate the cluster/smoothness and low-density assumptions of SSL. In those cases, adding more unlabeled data without necessary adaptation could worsen model performance \cite{van2020survey, amini2025self}. 

\section{Dataset}

This work built on a dataset curated in previous works \cite{chang2025multimodal, reverdy-etal-2022-roomreader}. Specifically, the current study further extended the processed dataset by incorporating new clips and multimodal features.

\subsection{Video clips}
\subsubsection{Videoconference corpus} 
We sourced the videoconference clips from the RoomReader corpus \cite{reverdy-etal-2022-roomreader}, which includes 30 Zoom sessions featuring 4 or 5 participants each, with durations ranging from 8 to 30 minutes. In total, the corpus comprises approximately 9 hours of recordings and 118 participants, 91 of whom are native English speakers. The conversations were primarily collaborative quiz games and icebreaker activities (e.g., "Family Feud"). The study was designed to encourage natural, spontaneous conversation, allowing participants to self-manage turn-taking organically.

\begin{figure}
\centerline{\includegraphics[width=0.5\textwidth]{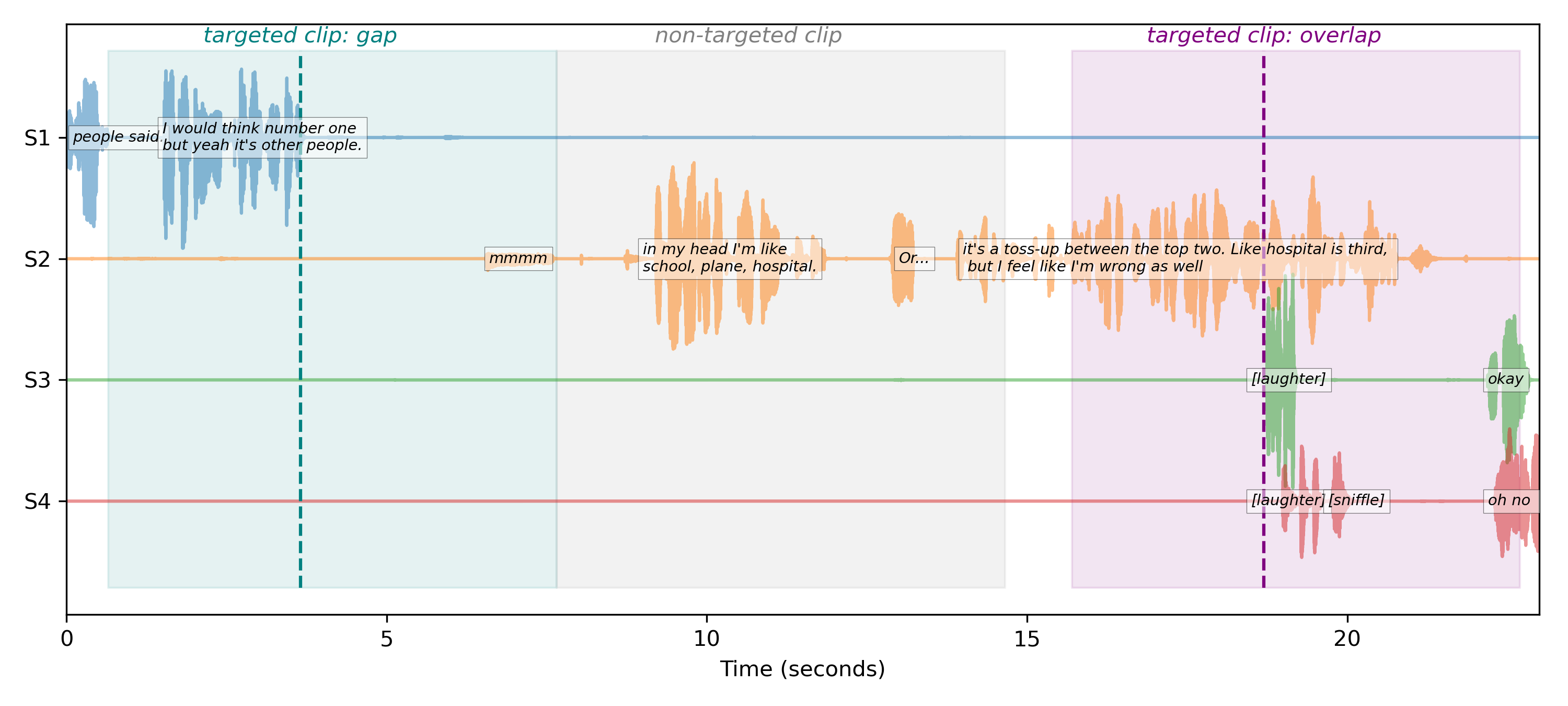}}
\caption{An example of the audio of natural turn-taking in a 4-people group conversation. We marked the start of a gap or overlap and extracted a 7-second "targeted clip" around each. These clips were then evaluated by multiple human annotators to identify those with low Fluidity or Enjoyment. The "non-targeted clips" were extracted from the remaining segments.}
\label{turn-taking}
\end{figure}

\subsubsection{Clip selection} 
As we are interested in the transient moments of negative experience, we clipped the continuous videoconference recordings into 7 second clips in both \textit{targeted} and \textit{non-targeted} approaches (Figure \ref{turn-taking}). 

\textit{Targeted clips}: We used speakers' audio to pinpoint moments likely linked to negative experiences, such as overlapping speech or unusually long gaps—instances. These indicate possibly disrupted turn-taking and were selected for targeted annotation in subsequent steps: (1) A root-mean-square (RMS) threshold of 0.05 (based on visual inspection of audio waveforms) was applied to detect individual speaker activity. Time points were marked as a gap when all speakers' audio remained below this threshold for at least 0.75 seconds—exceeding typical turn-taking gaps of 0.2 seconds in-person and 0.5 seconds in videoconferencing \cite{boland2022zoom}. Additionally, time points were marked when more than one speaker simultaneously exceeded the RMS threshold, indicating overlapping speech. (2) For each marked time point, a 7-second clip was extracted, spanning 3 seconds before and 4 seconds after the time point. There were a total 1508 targeted clips extracted for each type.

\textit{Non-targeted clips}: After extracting the targeted clips, we extracted all possible 7-second clips from the remaining segments (excluding the first and last 10 seconds), resulting in a total of 1340 non-targeted clips.

\subsection{Clip annotation}
\subsubsection{Annotators}
The annotators were New York University students who participated in the study and received course credit for their involvement. A total of 528 annotators (167 men, 274 women, 87 non-reporting or other; age 18-36), including 242 native English speakers, completed the survey. Annotators reported a median of 4 hours of video chat usage per week, distributed across interactions with friends and family (60\%), study or work-related activities (12\%), and classes or conferences (18\%), reflecting a mix of informal and formal contexts.

\subsubsection{Procedure}
Only the \textit{targeted clips} were annotated, with annotators unaware of the clip selection criteria. A red border highlighted the marked moment for 0.5 seconds. Annotations were completed on Qualtrics using personal computers and headphones in a quiet environment, and each rated 120 clips with the option to pause and resume within a week. To assess reliability, 8 identical targeted clips were shown at the start and end of the survey, while the remaining 104 clips were randomly sampled from all remaining targeted clips for each participant. For each clip, annotators rated it on a 5-point Likert scale (1: low, 5: high) based on the following questions: (1) "How \textit{fluid} is this conversation after the red frame appears?" (2) "How much did the group \textit{enjoy} the conversation?".

\subsubsection{Data preprocessing, reliability and inclusion}

As some annotators in online surveys tend to submit quick, inattentive responses, we used reliability trials to filter out these annotators' data. We assessed each annotator by calculating Pearson’s \textit{r} between their Fluidity ratings on 16 designated reliability clips and the average ratings from other participants. Annotators with an \textit{r} value above 0.2 were considered reliable, resulting in a pool of 350 annotators for further analysis. Clips that received fewer than four annotators were excluded, leaving 2,992 clips for analysis.

\subsubsection{Binary data transformation}

Since our primary focus is on detecting the most severe conversational failures, we transformed the Likert scale ratings into binary outcomes for machine learning modeling. Identifying low-rated clips was the main objective, making distinctions among highly and moderately rated clips (e.g., a rating of 3 or higher) less critical for this analysis. Although annotators provided ratings on a Likert scale, this approach was chosen to assess interrater reliability and better understand how the scale was utilized, rather than to collect binary judgments directly. A threshold of 2.5 was applied to both the Fluidity and Enjoyment scales, as this value corresponded to a noticeable gap in the data distribution observed in previous studies \cite{chang2025multimodal}. After binarization, the dataset included 2,731 clips rated high on both Fluidity and Enjoyment, 92 clips rated low on both, 123 clips with high Enjoyment but low Fluidity, and 46 clips with low Enjoyment but high Fluidity. A contingency analysis showed that while Enjoyment and Fluidity ratings were related, they were not identical (\textit{$\chi^2$} = 758.13, \textit{p} $<$ 0.001).

\subsection{Feature extraction}

\subsubsection{Audio embeddings}
Audio embeddings were extracted using VGGish \cite{vggish}, a widely used domain-general CNN-based audio DNN that processes spectrogram inputs and is pretrained on the YouTube dataset. Group audio recordings were first downsampled to 16 kHz, and VGGish then generated 128-dimensional embeddings for each 0.96-second segment. Although VGGish is no longer the state-of-the-art audio DNN, the model performances based on its feature embeddings were better than the alternative DNNs or transformers we have been explored (e.g., YAMNet, Wav2Vec2; not reported here). 

\subsubsection{Facial action embeddings}
Facial embeddings were obtained using the \textit{OpenFace 2.2} \cite{tadas2018openface}, which is a popular pretrained computer vision model for analyzing facial behavior in videos. It was applied on each videoconference participant's individual recordings to extract the intensity timeseries of 17 action units (e.g., blink, brow lowerer, but without lip suck). Note that we did not use the data provided by RoomReader dataset \cite{reverdy-etal-2022-roomreader}, which was likely generated by an earlier version of \textit{OpenFace}. 

\subsubsection{Dialogue text embeddings}
The dialogue text feature embeddings were extracted using \textit{all-MiniLM-L6-v2}. It is a compact and efficient model pretrained using the sentence transformers (SBERT) methodology, a variation of the BERT architecture that incorporates siamese and triplet network structures to generate sentence embeddings with rich semantic meaning \cite{reimers-2019-sentence-bert}. This model generates 384-dimensional sentence embeddings for tasks involving semantic similarity and clustering. The transcription data of RoomReader was retrieved for each clip, containing speaker, text, and timestamp details. The input text is then sorted chronologically based on start and end times, with each utterance formatted as “Speaker X: [text]” to preserve the dialogue’s structure. For example, utterances such as “Speaker A: Hello, everyone.” and “Speaker B: Hi! How are you today?” are used to construct coherent inputs.

\section{Modeling}

\subsection{Data split}
To analyze the impact of the labeled data ratio on SSL model performance, we partitioned the \textit{targeted clips} data into 10 folds. We employed \textit{StratifiedGroupKFold()} from \textit{sklearn} to preserve the approximate distribution of high- and low-rated clips while ensuring that data from the same videoconference session remained within the same fold. This approach enhances the model's ability to generalize to new videoconference sessions with different participants. Each fold contains approximately 300 clips from 3 sessions.

We employed a cross-validation strategy to train and evaluate the performance of our SSL models. In each iteration, 2 folds were designated as the holdout test set. From the remaining 8 folds, between 1 and 8 folds were used as labeled data, while the remaining clips, including the \textit{non-targeted clips}, were treated as unlabeled data. Each training fold represented approximately $\sim$8\% of the total dataset (comprising both labeled and unlabeled data in the training set). This process was repeated to exhaust all possible 11,475 combinations of train-test splits across the varying amounts of labeled training data.

\begin{figure}
\centerline{\includegraphics[width=.5\textwidth]{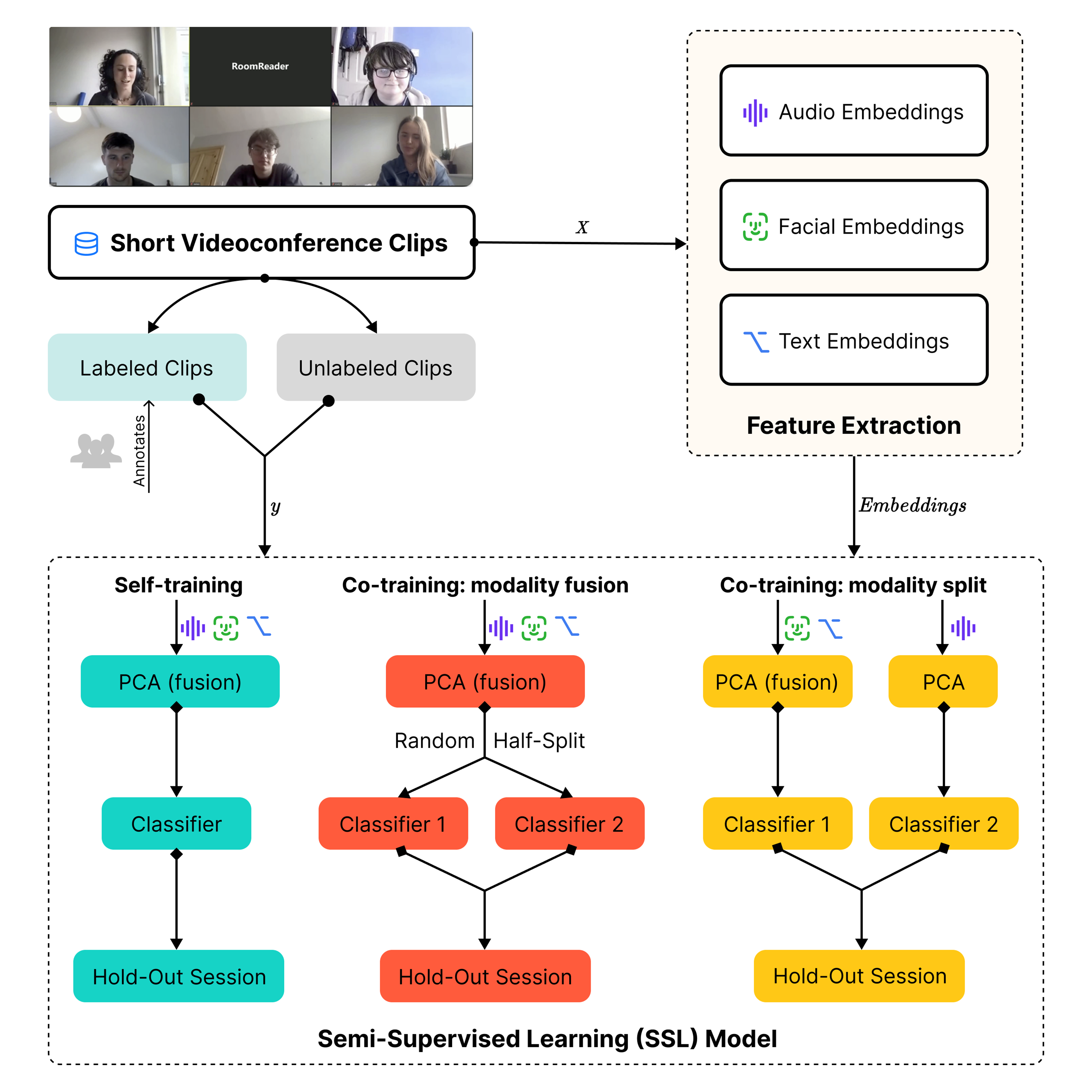}}
\caption{Multimodal feature extraction, modality fusion, and three SSL approaches.}
\label{diagram}
\end{figure}

\subsection{Model training}
We implemented SSL modeling using three approaches: (1) \textit{self-training}, (2) \textit{modality-split co-training}, and (3) \textit{modality-fused co-training} (Figure \ref{diagram}). These wrapper methods can be seamlessly combined with most of the SL base classifiers, and their flexibility and simplicity make them robust, reliable, and an increasingly popular initial choice for SSL. 

In the self-training algorithm, a base classifier iteratively assigns pseudo-labels to unlabeled datapoints whose confidence scores exceed a predefined threshold, thereby expanding the labeled dataset and retraining the base classifier \cite{amini2025self, yarowsky1995unsupervised}. 

Co-training is an extension of self-training that uses two base classifiers, trained on different feature subsets, to jointly assign pseudo-labels to unlabeled datapoints, offering potentially more robust performance \cite{van2020survey, wang2010cotrain}. We further implemented feature splitting in two alternative ways: (1) modality-split: audio vs. face+text features (to maintain a relatively balanced number of features), and (2) modality-fused: random splitting on PCA-transformed multimodal-fused features. The first approach assumes that different modalities provide independent views and can jointly assign pseudo-labels, while the later approach assumes that the emergent property in the fused multimodal data is necessary for the current task.

We implemented self-training and co-training using the \textit{scikit-learn} and \textit{sslearn} \cite{sslearn2025garrido} libraries, respectively. Prior to feeding input features into the model in self-training and modality-fused co-training, PCA was applied to all multimodal features. All PCs were fed into the self-training model as inputs, while in modality-fused co-training, they were randomly split in half for the two classifiers. In modality-split co-training, a PCA was applied individually to the audio or face+text feature set. We trained a logistic regression model using stochastic gradient descent method, \textit{SGDClassifier()}, with balanced class weights. We used \textit{Optuna} \cite{akiba2019optuna} framework to do hyperparameter tuning with a tree-structured parzen estimator sampling algorithm (PCA explained variance: 0-- or 20--100\%; loss function: "log\_loss" or "modified\_huber"; penalty: L1 or L2, alpha: 1e-5--1e-2 (log scale); self-training criterion: "threshold" or "k\_best"; pseudo-label assignment threshold: 0--1)

To assess whether the SSL models truly benefit from the unlabeled data, we trained an SL counterpart of the model using the same base classifier setup for each combination of data split.

Note that, instead of using a neural network architecture, we opted for logistic regression due to its robustness with limited data. This aligns with the study's goal of investigating how to achieve strong performance with minimal labeled data.

\section{Results}

\begin{figure}
\centerline{\includegraphics[width=0.5\textwidth]{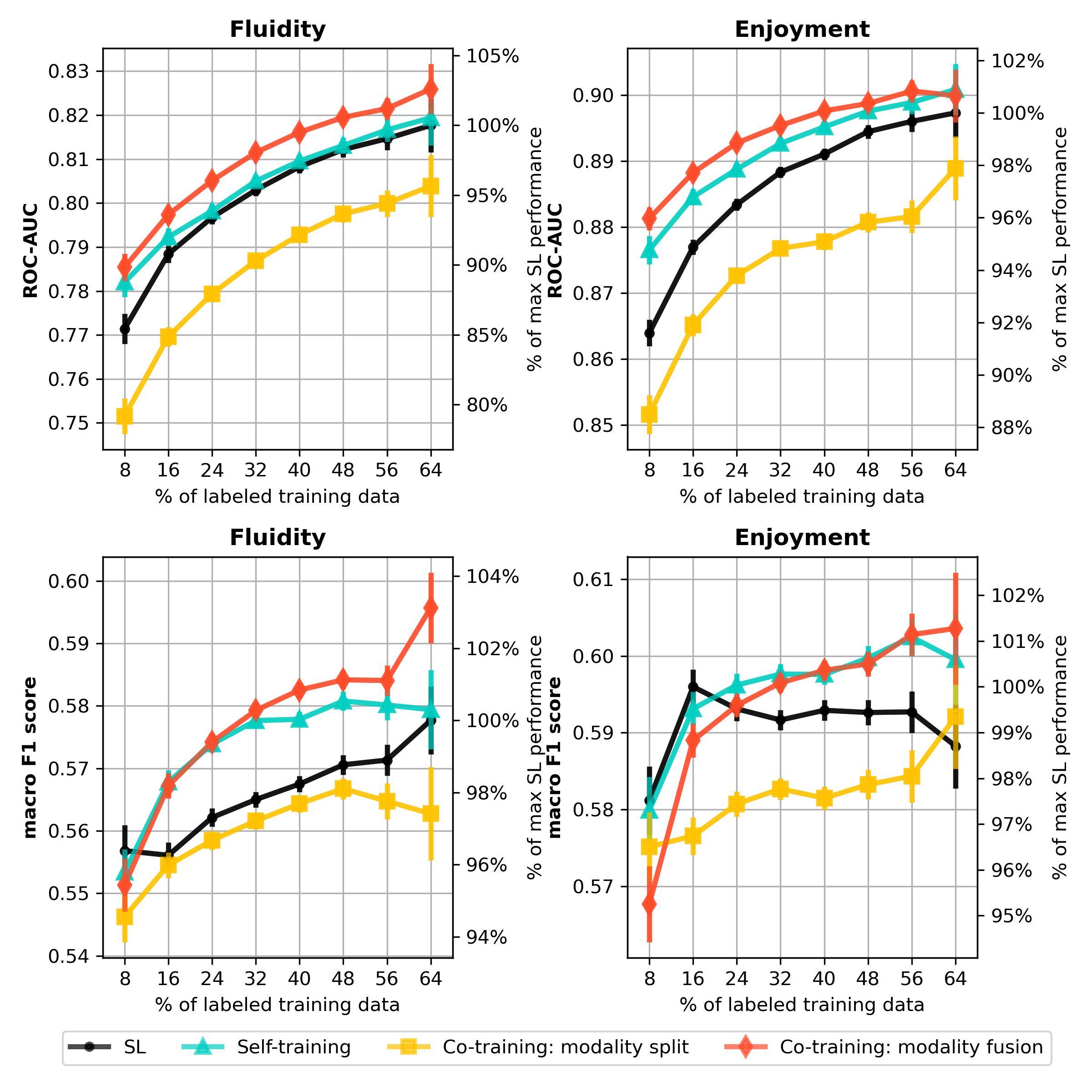}}
\caption{Performance of SSL and SL models. The modality-fused co-training SSL model outperforms SL models by up to 4\% with the same amount of labeled data, and using just 8\% of labeled data, it matches 95\% of the SL model’s full-data performance. (errorbar: standard error)}
\label{SSL_SL}
\end{figure}

Overall, modality-fused co-training model performed the best, followed by self-training (Figure \ref{SSL_SL}). They both outperformed SL counterparts at nearly every levels of labeled data in both ROC-AUC and macro F1 score by 1-4\% for predicting Enjoyment or Fluidity. Notably, when only 8\% of labeled data available, the ROC-AUC of the modality-fused co-training model can achieve 96\% performance of SL model’s full-data performance on predicting Enjoyment. Also, the modality-fused co-training model with as little as 24--32\% of labeled data can match or outperform the SL model’s full-data performance in both F1 and ROC-AUC on Enjoyment or Fluidity.

The benefit of SSL over SL can differ depending on the performance metric: SSL most significantly improves ROC-AUC over SL with only 8\% labeled data, F1 score benefits from SSL are less immediate, requiring at least 16-24\% labeled data for noticeable improvement. This lag likely stems from F1 score's sensitivity to class imbalance, and which effect makes unstable estimations with minimal labeled data.

\begin{figure}
\centerline{\includegraphics[width=0.5\textwidth]{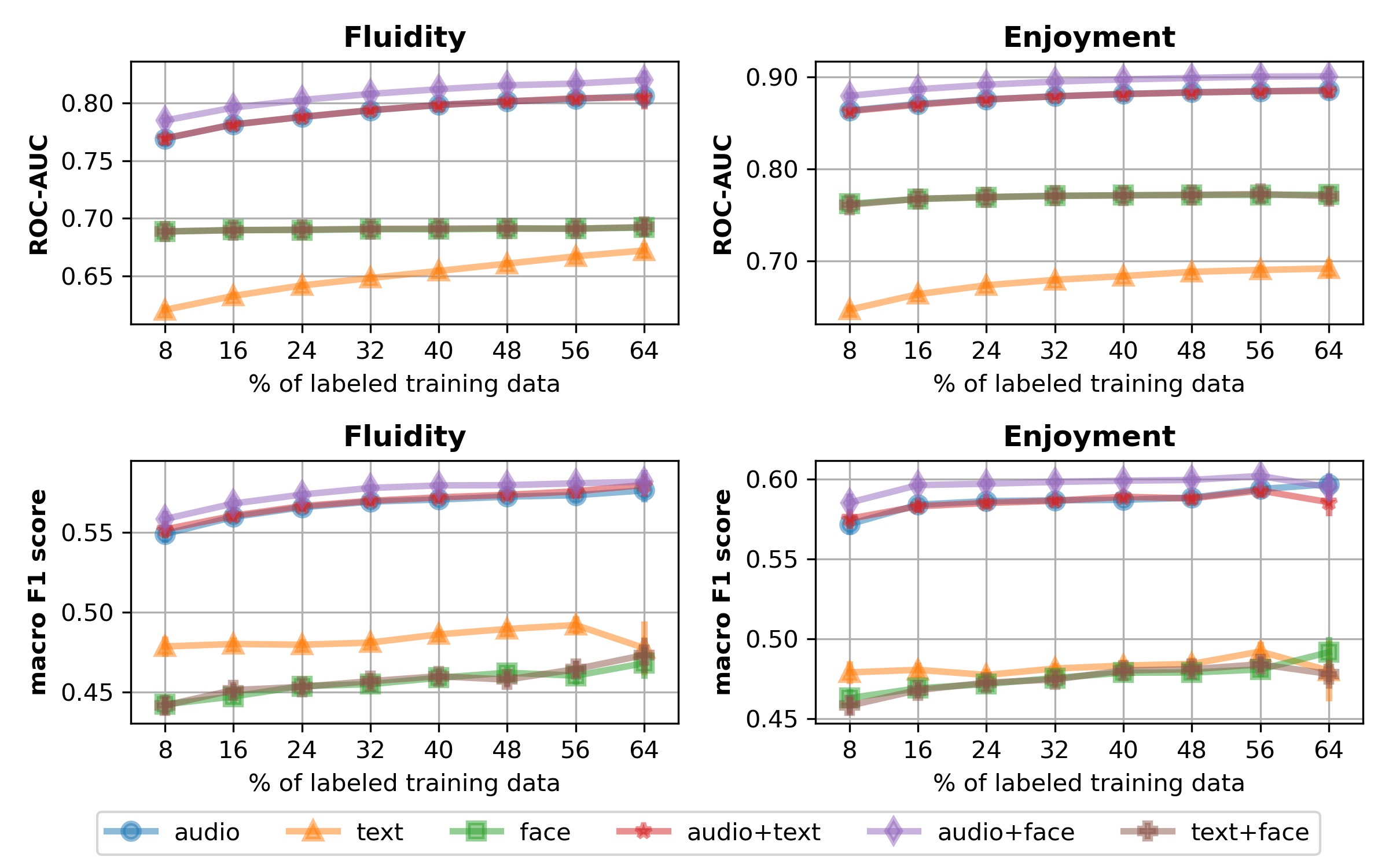}}
\caption{Ablation study: the fusion of the audio and face features is most critical for SSL modeling conversational experience in videoconferencing. (errorbar: standard error)}
\label{SSL_ind}
\end{figure}

We further conducted an ablation study to understand which feature modalities or their combinations are most informative by applying the self-training framework on every combination (not doing co-training here as the number of features can be too small to split) (Figure \ref{SSL_ind}). We found that the audio modality is most essential for the tasks, and its fusion with face modality can further improve the performance. Surprisingly, the text modality provide relatively little contribution. This implies the importance of the paralinguistic speech features, which can better be captured by an audio model than a BERT model, for the current tasks.

\section{Discussion}
This study demonstrated the efficiency of selectively targeting a few moments in continuously recorded data with minimal annotation to train a robust SSL classifier for detecting moments of negative experience. We also demonstrated this approach generalizes to new sessions with different participants. Even uneventful moments (i.e., non-targeted clips) can be leveraged within the SSL framework to further enhance model performance.

The modality-fused co-training approach performed best, likely for two key reasons: (1) The emergent property of modality-fused features is crucial for modeling conversational experiences. This is evidenced by the fact that the same co-training method performed well with modality-fused data but not with modality-split data. For example, the combination of text like 'haha' or 'oh no' with a quick, staccato tone versus a slow, flat tone can lead to very different semantic and pragmatic interpretations, showing emergent property can offer valuable and efficient insights for identifying moments of negative experience. (2) The co-training algorithm offers more robust generalization than self-training, with the benefit becoming more significant as the proportion of labeled data increases.

Notably, modality-split co-training consistently performed significantly worse than SL. While the underlying reason remains to be explored, this highlights that SSL algorithms do not always improve model performance and that the choice of approach and feature handling can have a significant impact.

While the multimodal transformer architectures are highly effective, achieving state-of-the-art performance in modeling speech emotion and turn-taking behaviors \cite{fatan20243m, li2021predicting, curto2021dyadformer, wagner2023dawn, lian2021ctnet}, the subjective aspects of conversational experience and videoconferencing remain largely unexplored — likely due to the lack of annotated datasets. Our approach addresses this critical gap by leveraging unlabeled data, paving the way for neural network models that can deeply understand conversational experiences and enhance videoconferencing interactions.

\section{Acknowledgements}
A.C., D.P., and D.F. are supported by NYU Discovery Research Fund for Human Health. A.C. is supported by National Research Service Award, NIDCD/NIH (F32DC018205) and Leon Levy Scholarships in Neuroscience, Leon Levy Foundation and New York Academy of Sciences. The funders have no role in study design, data collection and analysis, decision to publish, or preparation of the manuscript. This work was supported in part through the NYU IT High Performance Computing resources, services, and staff expertise.

\bibliographystyle{IEEEtran}
\bibliography{mybib}

\end{document}